\documentclass[12pt]{article}

\usepackage[colorlinks,linkcolor=blue,citecolor=blue,urlcolor=blue,bookmarks,bookmarksnumbered]{hyperref}
\usepackage{amsmath,amssymb,amsfonts}
\usepackage{slashed,extarrows,bigints,cancel}
\usepackage[paper=letterpaper,margin=1.0in]{geometry}
\usepackage{graphicx,xcolor}

\usepackage{cite}

\newcommand{\be}{\begin{equation}}
\newcommand{\ee}{\end{equation}}
\newcommand{\bea}{\begin{eqnarray}}
\newcommand{\eea}{\end{eqnarray}}
\newcommand{\ba}{\begin{array}}
\newcommand{\ea}{\end{array}}
\newcommand{\ben}{\begin{enumerate}}
\newcommand{\een}{\end{enumerate}}
\newcommand{\bi}{\begin{itemize}}
\newcommand{\ei}{\end{itemize}}
\newcommand{\bc}{\begin{center}}
\newcommand{\ec}{\end{center}}
\newcommand{\bfig}{\begin{figure}}
\newcommand{\efig}{\end{figure}}
\newcommand{\bq}{\begin{quotation}}
\newcommand{\eq}{\end{quotation}}
\newcommand{\bt}{\begin{table}}
\newcommand{\et}{\end{table}}
\newcommand{\btab}{\begin{tabular}}
\newcommand{\etab}{\end{tabular}}
\newcommand{\bs}{\begin{slide}}
\newcommand{\es}{\end{slide}}

\newcommand{\IR}{\mathbb{R}}

\newcommand{\beq}{\begin{eqnarray}}
\newcommand{\eeq}{\end{eqnarray}}
\newcommand{\beqn}{\begin{eqnarray}}
\newcommand{\eeqn}{\end{eqnarray}}

\newcommand{\rd}{\mathrm{d}}

\let\ba=\overline

\def\rd{{\rm d}}

\def\IR{\relax\leavevmode{\rm I\kern-.18em R}}
\def\ZZ{\relax\leavevmode
       \ifmmode\mathchoice
       {\hbox{\sf Z\kern-.4em Z}}
       {\hbox{\sf Z\kern-.4em Z}}
       {\lower.9pt\hbox{\scriptsize\sf Z\kern-.36em Z}}
       {\lower1.2pt\hbox{\tiny\sf Z\kern-.36em Z}}
       \else{\sf Z\kern-.4em Z}\fi}
\newcommand{\RR}{\mathbb{R}}

\def\resetby#1#2{\@addtoreset{#2}{#1}}
\def\seceq{\@addtoreset{equation}{section}
              \def\theequation{\thesection.\arabic{equation}}}

\def\Label#1{\label{#1}%
                \smash{\hbox to0pt{\raise1ex\hbox{\tiny[#1]}\hss}}}
\def\noLabels{\let\Label=\label}

\def\tx{\tilde{x}}
\def\m{\mathfrak{m}}
\def\tp{\tilde{p}}

\begin{document}

{\footnotesize
${}$
}

\bc

\vskip 1.0cm
\centerline{\Large \bf Quantum Gravity Phenomenology in the Infrared}
\vskip 0.5cm
\vskip 1.0cm

\renewcommand{\thefootnote}{\fnsymbol{footnote}}

\centerline{{\bf
Laurent Freidel${}^{1}$\footnote{\tt lfreidel@perimeterinstitute.ca},
Jerzy Kowalski-Glikman${}^{2,3}$\footnote{\tt jerzy.kowalski-glikman@uwr.edu.pl},
Robert G. Leigh${}^{4}$\footnote{\tt rgleigh@illinois.edu}
and
Djordje Minic${}^{5}$\footnote{\tt dminic@vt.edu  (Corresponding Author)}
}}

\vskip 0.5cm

{\it
${}^1$ Perimeter Institute for Theoretical Physics, 31 Caroline St. N., Waterloo ON, Canada\\
${}^2$ Institute for Theoretical Physics, University of Wroclaw,\\ Pl. Maksa Borna 9, 50-204 Wroclaw, Poland\\
${}^3$National Centre for Nuclear Research, Pasteura 7, 02-093 Warsaw, Poland \\
${}^4$ Illinois Center for Advanced Studies of the Universe \& Department of Physics,\\
			University of Illinois, 1110 West Green St., Urbana IL 61801, U.S.A.\\
${}^5$Department  of Physics, Virginia Tech, Blacksburg, VA 24061, U.S.A. \\
${}$ \\
}

\ec

\vskip 1.0cm

\begin{abstract}
Quantum gravity effects are traditionally tied to short distances and high energies.
In this essay we argue that, perhaps surprisingly, quantum gravity may have important consequences
for the phenomenology of the infrared.
We center our discussion around a conception of quantum gravity involving a notion of quantum spacetime that arises in metastring theory. This theory allows for an evolution of a cosmological Universe in which string-dual degrees of freedom decouple as the Universe ages. Importantly such an implementation of quantum gravity allows for the inclusion of a fundamental length scale without introducing the fundamental breaking of  Lorentz symmetry.
The mechanism seems to have potential for an entirely novel source for dark matter/energy. The simplest observational consequences of this scenario may very well be residual infrared modifications that emerge through the evolution of the Universe.

\end{abstract}

\vspace{1cm}


\begin{center}This paper has been awarded the Second Prize in the 2021 Essay Competition of the Gravity Research Foundation.
\end{center}

\renewcommand{\thefootnote}{\arabic{footnote}}

\newpage

Traditionally, searches for new physics, including quantum gravity, have been confined to high energies (UV) and small distances.
However, since the discovery of dark energy \cite{Riess:1998cb, Perlmutter:1998np}, the domain of low energies (IR) and long distances has become a frontier in fundamental physics. Recent searches for dark matter have
begun to concentrate on the very light and very weakly interacting degrees of freedom,
such as axions and axion-like particles \cite{Aprile:2020tmw} (for reviews and references consult \cite{Kim:2008hd,Graham:2015ouw}), as well as the so-called dark sectors, consisting of new light weakly-coupled particles that do not interact through the known strong, weak, or electromagnetic forces \cite{Essig:2013lka}.
Moreover, it is becoming increasingly clear at the fundamental level through the study of celestial holography \cite{Strominger:2017zoo,Kapec:2017tkm}, that the traditional description of particle physics is missing  an entire infrared sector of field theory.
Thus the exploration of deep infrared physics is becoming very important in the quest for an understanding of dark energy, dark matter and fundamental physics.

Similarly, motivated by the recent discovery of gravitational waves by LIGO \cite{Abbott:2016blz},
a new field of multi-messenger astronomy has arisen, which includes quantum gravity
phenomenology \cite{AmelinoCamelia:2002vw}.
In this context, a prominent proposal for exploring phenomenological implications of quantum gravity is
the search for UV modifications of dispersion relations, as well as other fundamental physical
relations, with the deformation scale commonly
taken to be of the order of the Planck scale \cite{AmelinoCamelia:2008qg, AmelinoCamelia:2009zzb,Mercati:2010au}.
However, in view of the relevance of infrared physics in the dark sector, infrared features of quantum gravity
should be explored as well.

In this essay we explore such aspects of quantum gravity phenomenology in the infrared.
We center our discussion on the concept of \emph{metaparticles} that we have introduced in previous work \cite{Freidel:2018apz}. Metaparticles refer to the zero mode sector of metastring theory, a reformulation of string theory in which T-duality plays a central role. As we will review below, the metastring theory gives rise to a novel realization of quantum spacetime that we call modular spacetime, and introduces new scales consistent with Lorentz covariance. In ordinary string theory, spacetime plays the role of an arena for particle dynamics, with the basic observables interpreted in terms of a particle S-matrix for the modes of the string. In the metastring theory on the other hand, such a spacetime only appears through decoupling, in the large scale limit. In this essay we describe a novel mechanism in which cosmological evolution is seen as a process within the larger context of modular spacetime.
In this context, there is a new parameter that appears in the kinematics of metaparticles that plays the role of an IR scale, and we argue, within a simple cosmological model, that as the Universe evolves, dual degrees of freedom increasingly decouple and may play the role of dark energy/matter. In the future, these ideas may lead to novel infrared quantum gravitational phenomena that might be searched for observationally.

Modular spacetime gives rise to a notion of quantum spacetime, corresponding to a generic polarization of quantum theory  \cite{Freidel:2016pls}. One way of thinking about metaparticles is that they can be obtained as the excitations of such a system.
On the other hand, metaparticles can be thought of as the generalization of particle excitations that appear in metastring theory \cite{Freidel:2015pka,Freidel:2017xsi,Freidel:2017wst,Freidel:2017nhg}. In both points of view, the fields that represent metaparticles are called \emph{modular fields}, whose properties depend on a fundamental length scale $\lambda$ while at the same time preserving Lorentz covariance.
The fact that the modular fields and their metaparticle excitations manage to resolve this fundamental conundrum of quantum gravity \cite{Hossenfelder:2012jw} is one of the main reasons that we believe they are of central importance in quantum gravity.

The modular fields which carry the metaparticle excitations are not simply functions of spacetime coordinates $\phi(x)$ but functions $\phi(x,\tilde{x})$, where $x^\mu$ are the usual spacetime coordinates while $\tilde{x}_\nu$ are dual coordinates conjugate to $x$. These coordinates are the eigenvalues of operators which do not commute,
\beq\label{xtxcomm}
[\hat{x}^\mu,\hat{\tx}_\nu]= i \pi \lambda^2  \delta^\mu_\nu \hat I.
\eeq
This commutation relation arose through a careful analysis of the zero mode sector of string theory  \cite{Freidel:2017wst}, and the scale $\lambda$ which appears in this commutation relation is for us a new phenomenological scale. In string theory, it happens to be the string scale $\lambda^2=\hbar\alpha'$ \cite{Freidel:2015pka}.
The key idea of modular spacetime is the fact that it is possible to diagonalize
the operators $(\hat{x},\hat{\tx})$ as long as we restrict \cite{aharonov2008quantum,Freidel:2016pls} their eigenvalues to lie\footnote{ The Heisenberg uncertainty principle  simply states that we cannot know  which specific modular cell inside phase space is chosen. }  inside a modular cell of size $\lambda$.
A central element of our proposal is the fact that the modular cell can be rescaled in a way that respects the commutation relation through
\be
x\to a x
\qquad \tx \to a^{-1} \tx .
\ee
In the limit of large expansion $a \to \infty$,  the dual dimension shrinks to zero and one recovers the usual spacetime notions. This is the \emph{decoupling} limit where a modular field restricts to the usual notion of a spacetime field.

To define the metaparticle one must introduce a notion of geometry adapted to the modular geometry. This  geometric structure, whose metric elements appear in the context of double field theory \cite{Hull:2009mi,Aldazabal:2013sca},
is called Born geometry \cite{Freidel:2013zga, Freidel:2014qna} when one also takes into account the symplectic structure of \eqref{xtxcomm}.
The simplest examples involve a metric $g_{\mu\nu} $ on spacetime, its inverse $ g^{\mu\nu}$ and a duality pairing
\be\label{metric}
\rd s^2 = g_{\mu\nu} \rd x^\mu \rd x^\nu +   g^{\mu\nu} \rd \tx^\mu \rd \tx^\nu,\qquad
\rd \tilde{s}^2 = \rd x^\mu \rd \tx_\mu.
\ee
The three elements of Born geometry which are the symplectic structure entering the commutation relation, the metric $g$ and the  duality pairing, unify symplectic, orthogonal and conformal geometries. The Born geometry is said to be flat when $g_{\mu\nu}$ is a flat metric. The rescaling of the modular cell can be reabsorbed as a
conformal rescaling $g_{\mu\nu} \to a^2 g_{\mu\nu}$ of the metric.

Born geometry is a necessary ingredient in the construction of metaparticles.
It will be sufficient for our purposes to consider free metaparticles, whose dynamics are given by a worldline action involving phase space coordinates corresponding to $x,\tilde x$. The worldline action is of the form
\cite{Freidel:2017wst, Freidel:2017nhg, Freidel:2018apz}
\begin{equation}\label{1}
S \equiv \int_0^1 d\tau \Big[p\cdot \dot x +\tilde p\cdot \dot{\tilde x}+ \pi \lambda^2 \, p \cdot\dot{\tilde p}
- {N}\left(\tfrac12 p^2 +\tfrac12 {\tilde p}^2 + \mathfrak{m}^2\right) +{\tilde N}\left(p\cdot \tilde p - \mu \right)\Big]\,.
\end{equation}
Here the signature is $(-,+,\ldots,+)$ and the contraction of indices defining the duality pairing is denoted by $\cdot$.
At the classical level, the theory has worldline reparameterization invariance,
which is due to the presence of the mass-shell constraint.
This model possesses  two additional features \cite{Freidel:2018apz}:
the first new feature of the model is the presence of an additional local symmetry which from the particle worldline point of view is associated with an additional local constraint. The second new feature is the presence of a non-trivial symplectic form on the metaparticle phase space, the non-zero brackets being
\beq
\{p_\mu, x^\nu\}=\delta_\mu^\nu,\qquad
\{\tilde p_\mu, \tilde x^\nu\}=\delta_\mu^\nu,\qquad
\{ \tilde x_\mu,x^\nu \}=  \lambda^2 \delta_\mu^\nu,
\label{xtxcomm1}
\eeqn
with $\mu,\nu=0,1,...,d-1$.  The last relation corresponds with \eqref{xtxcomm}.

The attractive features of this model include worldline causality and unitarity, as well as an explicit mixing of widely separated energy-momentum scales. To proceed with a Hilbert space description of metaparticle states, one must choose a polarization of the phase space, and the simplest example is to regard states as provisionally labeled by $p$ and $\tilde p$, as $|p,\tilde p\rangle$. We will typically interpret $p_\mu$ to represent ordinary spacetime momentum, with $\tilde p^\mu$ as additional 'internal' quantum numbers. This corresponds to a choice of how spacetime is interpreted to appear within Born geometry and can be thought of as a property of the decoupling limit.
 Physical  states will be annihilated by the constraints implied by \eqref{1},
\begin{align}
\tfrac12 p^2 +\tfrac12 {\tilde p}^2 + \mathfrak{m}^2 =0,\label{1const}\\
p\cdot \tilde p - \mu =0.\label{2const}
\end{align}
Although we will not consider metaparticle interactions directly in this paper, such interactions preserve\footnote{This means that metaparticle amplitudes are proportional to
$ \delta^{(d)}(\sum_i p_{i})\delta^{(d)}(\sum_i  \tilde p_{i})$. Note that we are suppressing any additional labels on states, such as spin or helicity. The structure of interactions is itself an interesting and largely unexplored subject, but one expects that the interactions are spin-dependent.} both $p$ and $\tilde p$.
The metaparticle propagator between off-shell states follows from the worldline path integral involving the above action
and it has the following form \cite{Freidel:2018apz}
\begin{equation}\label{doubletramp}
G(p,\tilde p; p_i,\tilde p_i)
\sim
\delta^{(d)}(p-p_{i})\delta^{(d)}(\tilde p-\tilde p_{i})
\frac{\delta(p\cdot\tilde p-\mu)}{p^2+\tilde p^2+2\mathfrak{m}^2-i\varepsilon}.
\end{equation}
We see that although $p,\tilde p$ might resemble a doubling of momentum space, the propagator contains a \emph{simple} pole along with a $\delta$-function constraint.
We conclude from this analysis that a generic metaparticle is characterized by three invariant scales, each associated with an element of the Born geometry:
the fundamental non-commutativity scale $\lambda$, the metaparticle mass scale  $\m$ and the duality scale $\mu$.

As mentioned above, the metaparticles represent the fundamental particle-like excitations of the metastring, which is a formulation of string theory where  T-duality symmetry is manifest \cite{Freidel:2015pka,Freidel:2017xsi,Freidel:2017wst,Freidel:2017nhg}.
In the usual interpretation of string theory, it is often said that string excitations can be interpreted as a collection of particles in spacetime with a hierarchy of evenly spaced masses. Such a statement is only true {\it after} one has taken the decoupling limit. To see this, consider the spectrum of string theory
compactified along  timelike and  spacelike directions with radius of compactification given by
$R = a \lambda$, where $\lambda$ is the string scale and $a$ a scale factor in conformally flat coordinates.
The compactified string spectrum is given by a collection of metaparticles with mass scale $\m$ and duality scale $\mu$ satisfying
\be\label{spec}
- \tfrac12 a^{-2} p^2 - \tfrac12 a^{2}{\tilde p}^2= \m^2 =\frac1{\lambda^2}(N_L+N_R-2),\qquad
p\cdot\tilde{p}=\mu= \frac1{\lambda^2}(N_L-N_R),
\ee
where we have also rescaled the non-compact components of $p$, and where $N_L$ (resp. $N_R$) is the left (resp. right) oscillator number.
$(p,\tp)$ denotes the  `comoving' momentum and dual momentum at the self dual radius,
while $(P,\tilde{P})=(a^{-1}p, a\tilde{p})$ are the physical momenta.
The usual string spectrum on flat spacetime corresponds to all states having $\mu=0$ and $\tilde p=0$. This is actually a superselection sector.
The choice of $\mu=0$ imposes the level matching condition, whereas flat compactifications correspond to including states with non-zero $\tilde p$, with the restriction that $\tilde p$ is non-zero along only space-like directions.
The decoupling limit $a \to \infty$ of the compactified string can only be achieved
as a singular limit, for states that satisfy $\tilde{p}^2=0$, hence $\tilde{p}=0$ and $\mu=0$.
The metaparticle sector takes on its full power once we lift these restrictions: one does not restrict  the dual momenta to  be purely spacelike and one does not impose the vanishing of the duality scale $\mu$.
It is clear from \eqref{spec} that the spectrum is invariant under the exchange $p \leftrightarrow \tp$ which corresponds to T-duality. Accordingly, when the scale factor goes to zero the dual decoupling limit means that a dual universe made of  dual particles is accessed. Although one may worry that these relaxations cannot be made consistent with causality, the form of the propagator \eqref{doubletramp} essentially resolves the issue \cite{Freidel:2018apz}.

We consider first a flat Born geometry $a=1$. The constraints (\ref{1const},\ref{2const}) are then together invariant under a generalization of the Poincar\'e group given by $(O(1,d-1)\ltimes \tilde{O}(1,d-1))\ltimes\RR^{2d}$. The orthogonal algebras act on the momenta as
\be
\delta_\alpha (p_\mu,\tp^\mu)  = (\alpha_{\mu}{}^{\nu} p_\nu, \tp^\nu\alpha_\nu{}^\mu),
\qquad
 \delta_{\tilde\alpha} (p_\mu,\tp^\mu)  = (\tilde\alpha_{\mu \nu} \tp^\nu,
 \tilde\alpha^{\mu \nu} p_\nu).
\ee
where $\alpha_{\mu\nu}=-\alpha_{\nu\mu}$ and $\tilde\alpha_{\mu\nu}=-\tilde\alpha_{\nu\mu}$
are infinitesimal Lorentz parameters. We refer to the first action as the Lorentz and the second as the dual Lorentz actions.
The Lorentz action acts diagonally on the momenta, and in particular acts in the usual way on the momentum $p$.
There are three  Lorentz invariants:
$\tfrac12(p^2+\tilde{p}^2)$,  $p\cdot\tilde p$ and $\delta : =\frac12(\tilde p^2-p^2)$.
The first two are also invariant under dual Lorentz transformations, and they form the two metaparticle constraints.
The last combination $\delta$ is not fixed but is modified
by
the action of  $\tilde{O}(1,d-1)$. This means that $\delta$ parameterizes the orbit of $\tilde{O}(1,d-1)$.

Now the puzzle we are facing is that the usual notion of mass depends on the value
of the orbit parameter $\delta$,
\be
m^2:=-p^2 = (\m^2+ \delta).
\ee
This means that the value of this mass depends on the choice of the dual frame of reference and can be modified by the action of a dual boost!
It is clear that if one wants to get phenomenologically acceptable predictions from a theory of metaparticles, we need a mechanism to fix the value of $\delta$ in the universe we live in today.

 To understand how to resolve this problem, one first needs to appreciate that the presence of the non-commutativity scale $\lambda$ means that the dual symmetry is fundamentally broken at the quantum level. This follows from the fact that the dual Lorentz symmetry maps $\hat{x}^\mu$ onto its conjugate variable and therefore  the commutativity of the spacetime coordinate is not preserved
 \be\label{braxx}
 \delta_\alpha([\hat{x}^\mu,\hat{x}^\nu])= 2\pi i\tilde\alpha^{\mu\nu}.
 \ee
 In other words the dual Lorentz action is not a canonical transformation.
 Now since the modular fields and their decoupling limit are obtained by a choice of polarization it is clear that they do not transform covariantly under the dual Lorentz symmetry.
 In other words the dual Lorentz symmetry that acts on the free one metaparticle state is broken.  The symmetry breaking happens at the level of the vacuum sector, and thus a choice of ground state fixes a value of $\delta$.

 To see this clearly we can look at the decoupling limit in which the fields become simply functions of $x^\mu$. It is of paramount importance that the spacetime coordinates which appear in the decoupling limit are \emph{commutative}. As shown by \eqref{braxx}
 this is not the case for  $x^\mu$ after a dual boost. Once a dual boost is performed the new commuting decoupling coordinate is no longer $x$, but $ x^\mu+\alpha^{\mu\nu}\tx_\nu +\cdots$,
 and the particle mass is the mass with respect to the momentum conjugate to this new decoupling coordinate.
 After a dual boost the spacetime coordinate has been rotated into the non-commutative realm and it is no longer acceptable to assume that $p_\mu$ is the momentum measured by asymptotic observers.

 We now propose that the expansion of the Universe can act as a mechanism that selects the
 decoupling frame and thus a value of $\delta$.
 Before doing so, we note that one can show that there is a classification of metaparticle states whose mass-squared is positive, $\m^2\geq 0$.
We  assume without loss of generality that $\mu \geq0$.
In fact there are two distinct massive classes corresponding to the values
\be
\mathrm{I}: \, \m^2 >\mu \qquad
\mathrm{II}:\, \mu>\m^2 \geq 0, \label{IandII}
\ee
and in addition the `massless' class $\m^2= \mu$. One can show that the metaparticles of class I are such that $\delta^2 > \m^4-\mu^2 $ while $\delta$ can take any value in class II.

 Since the physical states are labelled\footnote{Note that such states are naively in conflict with the Coleman-Mandula theorem \cite{Coleman:1967ad}. However, given that the dual Lorentz is broken and in consideration of the following discussion, we expect that this conflict can be resolved and in any case deserves careful consideration.} by pairs $(p,\tp)$ we will interpret  $p_\mu$  as the physical momentum, while $\tp^\mu$ can be regarded as some internal quantum numbers.
 We will suppose that $p^2< 0$ and we can then  use the first Lorentz action to
 go to the center of mass frame and set
\beq
p_\mu=(m,\vec 0),\qquad \tp^\mu=\left(\frac{\mu}{m},\tilde{\mathbf{p}} \right),\qquad \tilde{\mathbf{p}}^2
= \left( \frac{\mu}{m}- m\right)^2 +2(\mu-\m^2).
\eeq
We see that
for class II metaparticles we have
$\tilde{\mathbf{p}}^2\geq 2(\m^2-\mu)$ while for class I metaparticles we can always choose a dual frame
where $\tilde{\mathbf{p}}=0$. For such a state,
$
\delta=m^2-\m^2$.

We now make use of the fact that the metaparticle couples to the extended metric \eqref{metric}, so  in the case of a cosmological background
\begin{equation}\label{metr}
  g_{\mu\nu}dx^\mu dx^\nu = - dt^2 + a_c^2 d\mathbf{x}^2,
\end{equation}
and  for metaparticle momenta $p=(E,\mathbf{p})$ and
$\tilde{p}=(\tilde{E},\tilde{\mathbf{p}})$  we get that\footnote{ We assume that the cosmological scale factor $a_c$ evolves slowly and we take the instantaneous dispersion relation.}
\begin{equation}\label{cosmdisp}
E^2 - a_c^{-2} {\mathbf{p}^2} + \tilde{E}^2 - a_c^2 \tilde{\mathbf{p}}^2 = 2\m^2,
\qquad
E\tilde{E} + \mathbf{p}\cdot \tilde{\mathbf{p}}=\mu.
\end{equation}
Note that as the scale factor $a_c$ gets larger the dual momentum becomes smaller, and for
a Universe of small size $a_c$ the dual momentum is larger.
Even this very simple type of reasoning offers a new view on cosmology and the origin of the Universe,
which takes into account both the visible and dual degrees of freedom.
Cosmological models for the generalized metric \eqref{metric} have already been developed in   \cite{Brandenberger:2018bdc,Bernardo:2019pnq} and references therein.

We clearly see from the first equation of \eqref{cosmdisp} that the only states that  survive at late times when $a_c\to \infty$ are the states where the dual momentum $\tilde{\mathbf{p}}$ vanishes. This means that all metaparticles of class II decouple and that  the metaparticles of class I become particles with physical momenta
given by $P=(E, \mathbf{P})$ with $\mathbf{P} = a_c \mathbf{p}$ and dispersion relation
\be
E^2 +\frac{\mu^2}{E^2}-  \mathbf{P}^2 = 2 \m^2.\label{mudisp}
\ee
Since for class I metaparticles $\mu<\m^2$, we interpret the duality scale $\mu$ as an infrared scale.  It is clear that for the dispersion relation to be compatible with observations, the scale $\mu$ must be very small. For the lightest massive particles, neutrinos, the cosmological bound on the sum of neutrino masses is $\sum_i m_i = 0.097 eV$ \cite{pdg}. Assuming that $\mu_{\nu}/E^2$ is extremely small for MeV energies at which the neutrino massess are measured, we can safely assume that $\m_\nu \sim 10^{-1}\, eV$. It follows then from \eqref{IandII} that for neutrinos  the parameter $\mu_{\nu} \lesssim 10^{-2}\, eV^2$. Note that this characteristic energy is very close to the scale of dark energy. Another interesting aspect of this dispersion relation is that the crucial physics appears in the infrared as opposed to the ultraviolet regime, usually expected in the context of quantum gravity phenomenology \cite{AmelinoCamelia:2008qg}.

The dispersion relation \eqref{mudisp} is Lorentz violating.
What is interesting is that this Lorentz violation comes from a fundamental theory which is fully Lorentz covariant.
In fact this relation follows from the general covariant dispersion \eqref{cosmdisp} in the case of an old Universe and holds in the cosmological frame only.
The Lorentz violation is due to a coupling between the fundamental degrees of freedom needed to UV-complete the theory and the expansion of the Universe.
It simply comes from the fact that the expansion of the Universe gives rise to a preferred frame and the infrared Lorentz violation appears only for an old Universe, as the symmetry is restored as one goes back in time. The coupling, through the duality scale, between some fundamental elements of the metaparticle theory and the Universe's evolution is one of the most interesting features of our model.

In this essay, we have considered only the simplest implementation of cosmology in the context of string duality. One can imagine more elaborate scenarios in which the geometry \eqref{metric} is replaced by a solution with an evolving duality frame. In any case, the mechanism that we have described here seems to have potential for an entirely novel source for dark matter/energy, with states of class II being remnants, and introduces new notions of Universe evolution, with dual degrees of freedom being more relevant early on and decoupling at late time. The simplest observational consequences of this scenario may very well be the infrared modifications identified above.

\noindent
{\bf Acknowledgments:}
We thank Per Berglund, Patrick Huber, Tristan H\"{u}bsch and Tatsu Takeuchi for
discussions. DM and JKG thank  Perimeter  Institute  for  hospitality.
LF, RGL and DM thank  the  Julian  Schwinger Foundation  for  support.
RGL is supported in part by the U.S. Department of Energy contract DE-SC0015655 and DM by the U.S. Department of Energy under contract DE-SC0020262.
For JKG, this work was supported by funds provided by the National Science Center, project number 2017/27/B/ST2/01902 and 2019/33/B/ST2/00050.
Research at Perimeter Institute for Theoretical Physics is supported in part by the Government of Canada through NSERC and by the Province of Ontario through MRI. This work contributes to the European Union COST Action CA18108 {\it Quantum gravity phenomenology in the multi-messenger approach.}

\end{document}